\documentclass{article}

\usepackage[utf8]{inputenc}
\usepackage[T1]{fontenc}
\usepackage{url}
\usepackage{fullpage}
\usepackage{amsthm,amsmath,amssymb,microtype,graphicx,subcaption,booktabs,enumitem,hyperref,rotating,caption,calc}
\usepackage[capitalise]{cleveref}
\usepackage{tikz}
\usepackage[normalem]{ulem}
\usepackage{authblk}
\usepackage{wrapfig}

\DeclareCaptionStyle{figstyle}
  [format=plain,margin=0pt,justification=centering]
  {format=plain,margin=8pt}
\def\ShowAuthNotes{1}

\ifnum\ShowAuthNotes=1
\newcommand{\authnote}[3]{\textcolor{#3}{[{\footnotesize {\bf #1:} { {#2}}}]}}
\else
\newcommand{\authnote}[3]{}
\fi

\definecolor{DarkBlue}{RGB}{0,0,150}
\definecolor{DarkRed}{RGB}{150,0,0}
\definecolor{DarkGreen}{RGB}{0,150,0}

\newcommand{\Encrypt}{\mathsf{Encrypt}}

\newcommand{\Decrypt}{\mathsf{Decrypt}}

\newcommand{\sk}{\mathsf{sk}}

\newcommand{\sign}{\mathsf{sign}}
\newcommand{\verify}{\mathsf{verify}}
\newcommand{\hash}{\mathsf{hash}}
\newcommand{\hashtree}{\mathsf{hash\_tree}}
\begin{document}

\title{Safepaths: Vaccine Diary Protocol and Decentralized Vaccine Coordination System using a Privacy Preserving User Centric Experience}

\author[1]{Abhishek Singh}
\author[1]{Ramesh Raskar}
\author[2]{Anna Lysyanskaya}
\affil[1]{\footnotesize MIT Media Lab}
\affil[2]{\footnotesize Brown University}

\maketitle

\begin{abstract}
In this early draft, we present an end to end decentralized protocol for the secure and privacy preserving workflow of vaccination, vaccination status verification, and adverse reactions or symptoms reporting. The proposed system improves the efficiency, privacy, equity and effectiveness of the existing manual system while remaining inter-operable with its capabilities. We also discuss various security concerns and alternate methodologies based on the proposed protocols.
\end{abstract}

\section{Introduction}
\paragraph{Motivation:} Recent announcements of vaccines have created a sense of hope for the near future of the society currently burdened with lock-downs and quarantines. However, a lot of effort and coordination is required to go from successful vaccines to successful vaccination programs to curb the disease spread. We believe a user-centric design using vaccination cards and/or mobile phones can play a critical role in the micro-planning and last mile issues. In this work, we integrate  work in user privacy, cryptography, and user interaction to design secure and private protocols which span from the starting step of vaccination program enrolment to all the way to symptoms reporting from potential side effects of vaccination.\\
We consider the first four of the following parts of the system:
\begin{itemize}
    \item 
    Indicating eligibility based on priority tiers (anonymity via vaccine coupons), 
    \item
Second dose coordination (privacy preserving record linkage), 
\item
vaccine verification/passports (interoperability and privacy), 
\item
Safety and efficacy monitoring (crowdsourced monitoring of safety and efficacy using private aggregation)
\item
Trust and communication (social media analytics, contextual messaging)
\end{itemize}
\begin{figure}[h]
    \centering
    \includegraphics[width=0.9\columnwidth]{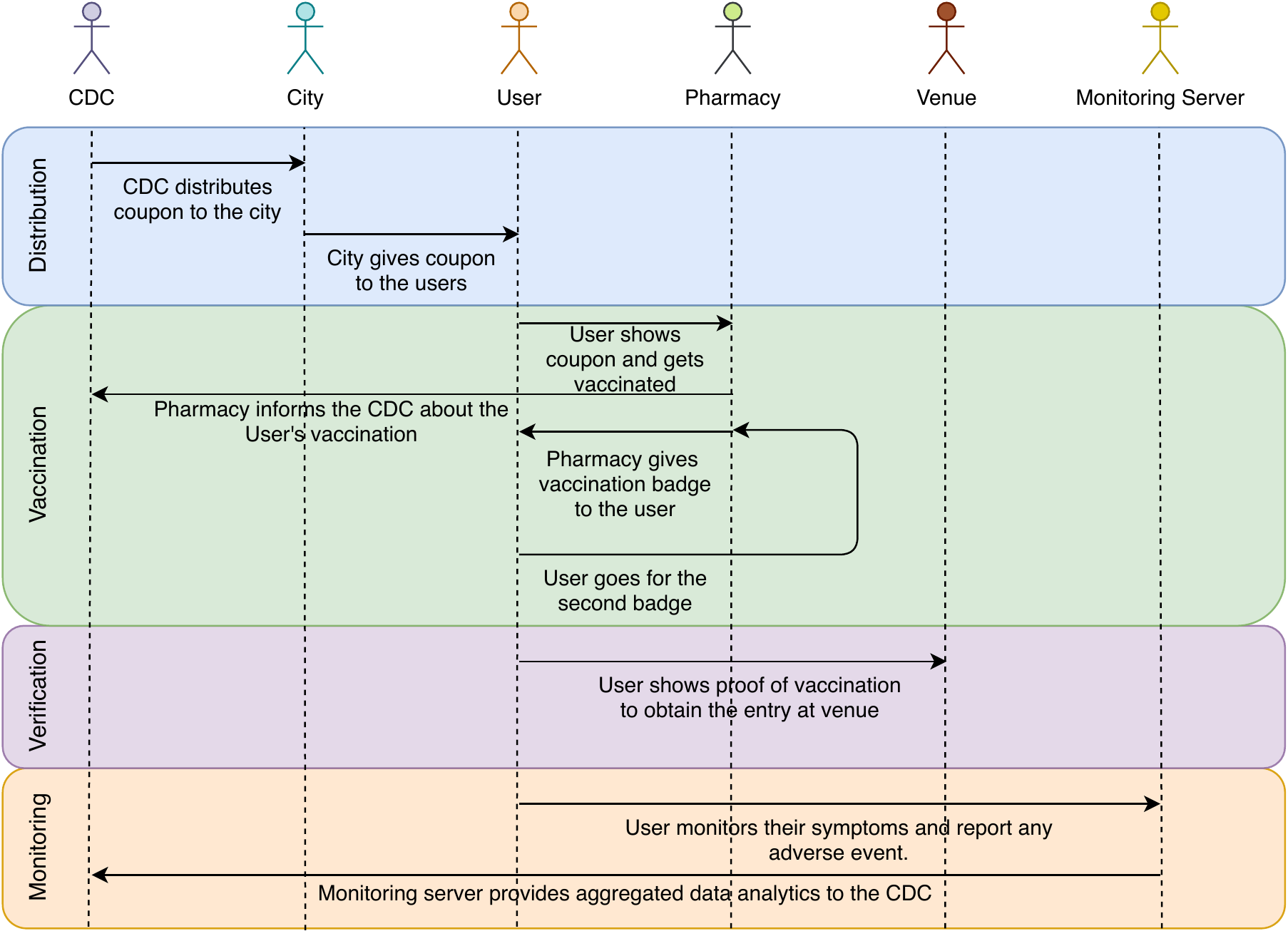}
    \caption{Entity and information flow depicting the system at a high level. We project this diagram as the backbone of the proposed system. We describe the detailed information flow and privacy preserving mechanisms in the next figures.}
    \label{fig:my_label}
\end{figure}

\paragraph{Participants:} Let us consider the different participants and their corresponding roles.
\begin{itemize}
    \item \textit{Issuer} is a trusted entity that initiates the enrolment process by distributing the coupons which would be eventually used for getting vaccinated as described later. In our use-case the \textit{issuer} could be CDC or any similar authoritative body that currently monitors the distribution of the vaccines.  Later, the issuer also issues the vaccination credentials, or badges.  Strictly speaking, the latter issuer (\textit{the badge issuer}) need not be the same entity.
    \item \textit{Distributor} is the entity that receives the vaccination coupons from the \textit{issuer} and distributes it locally to individuals. In our use-case the \textit{distributor} could be the city government office.
    \item \textit{User} is the person who wants to get vaccinated and use their vaccination status later on to obtain a vaccination passport that will be used as a proof of vaccination status.
    \item \textit{Pharmacy} is the entity that performs the vaccination. This could be nearby health/pharmacy stores like CVS.
    \item \textit{Verifier} refers to the set of authorized users that can verify the vaccination status of any \textit{user} after their consent. This can be a venue owner that is managing access to their facilities.
    \item \textit{Healthcare researcher} is a participant who may be on the receiving end of alerts from users who choose to provide feedback for the benefit of the community.
\end{itemize}

\paragraph{Phases:}  We model the vaccination process as consisting of the following phases:

\begin{itemize}
    \item \textit{Setup and registration phase} in which the entities establish a trust infrastructure.  In particular, the trusted Issuer generates its cryptographic keys and publishes the public half of those keys.  Other participants that require keys generate their keys as well and obtain certificates on the public parts of their keys.  For example, a distributor may want to be able to digitally sign a statement such as ``Alice is eligible for vaccination."  Additionally, pharmacies need a way to authentically communicate the fact that Alice has been vaccinated.
    \item \textit{Coupon distribution phase} in which the Issuer and Distributor jointly create, for each eligible user $U_i$ a string $c_i$ that constitutes the user's vaccination coupon, and communicate it to the user.
    \item \textit{Vaccination phase one} in which a User uses his coupon $c_i$ possibly together with other forms of identification (or biometrics) to convince a Pharmacy of his eligibility to receive the vaccine. If the pharmacy determines that the coupon is valid for this user, the user receives a first dose of the vaccine together with a badge $b_i$ that can be used as proof of vaccination.  
    \item \textit{Vaccination phase two} in which a User uses his badge $b_i$, possibly together with other forms of identification (or biometrics) to convince the Pharmacy that he's eligible for a second shot of the vaccine. If the pharmacy determines that the badge is valid for this user, the user receives the second shot of the vaccine together with an updated badge $b_i$ that can be used as proof of vaccination.  
    \item \textit{Verification phase} in which a User uses his badge $b_i$, possible together with other forms of identification or biometrics to convince a Verifier of his vaccination status.
    \item \textit{Gathering useful data phase} A user who has been vaccinated can use his badge $b_i$ to contact healthcare researchers in case of adverse effects or to report other helpful information about the efficacy of the vaccine.
\end{itemize}

\subsection{Security, privacy, and usability considerations}

We consider different aspects of the problem from a practical standpoint:

\begin{description}
\item[Unforgeability of coupons]  No adversary, even if he controls the distributor, should be capable of creating functional and valid coupons without say-so from the \textit{issuer}.  Moreover, a coupon cannot be used by more than one user.

\item[(Optional) Non-transferability of coupons] If both the \textit{issuer} and the \textit{distributor} are honest, then a coupon issued to user $i$ cannot be used by any other user $j$ where $j\neq i$.  This is optional as precise personalization of coupons requires information about specific eligible individuals, and this information may be unavailable or too costly to obtain when the coupons are created.  Moreover, users have an incentive not to transfer their coupons to others, so that they can get vaccinated themselves, so personalization may be unnecessary.

\item[Unforgeability and non-transferability of badges] Similar to the unforgeability and non-transferability properties of the coupon, the badges should also have the same property. This would mean that no entity should be able to create functional and valid coupons without say-so from either the \textit{distributor} or the \textit{issuer}.

\item[(Trade-off) Off-line phases]  A phase is off-line if it does not involve that any particular trusted participant who isn't explicitly part of the phase be online (require internet connection to exchange data with servers). The more phases are off-line, the easier are the logistics and equity aspects of it.

\item[User privacy from the issuer]  The issuer (assuming computationally bounded) should not be able to infer the identity of the vaccinated user; from the issuer's point of view there is no difference whether it was Alice or Bob. This property prevents against any centralization of the identity of the vaccinated users by any untrusted entity.

\item[(Trade-off) User privacy from the verifiers]  The verifier should not learn anything about a user other than her vaccination status after performing the verification protocol.

\item[(Trade-off) Non-traceability] The verifiers should not be able track the user across different venues even if they collude with each other.

\item[Accessibility]  Any user, even one who does not have access to smartphones or other possibly costly devices, should be able to participate in the protocol.

\item[Usability] The system should be easy, intuitive to use, not require excessive training to set up, not require a significant amount of time to use.  This applies not just to users, but also to whoever administers the other components of the system: the admins of the issuer, the distributors, the pharmacies, etc.
\end{description}

\textbf{Trade-offs:} The cost of having the vaccination phase offline is that each \textit{Pharmacy} must be able to issue a badge by itself, meaning that it will need to have a secret key, use digital signatures, etc; also either verification of a badge will require that the verifier know the pharmacy's public key, or that some more sophisticated cryptographic algorithm (such as delegatable anonymous credentials~\cite{cryptoeprint:2018:923}) be used.

Unless the badge is physically embedded into a user's body, it is hard to achieve non-transferability of badges and user privacy from verifiers at the same time.  The best we can do is to relax the assumption and have verifier learn some attribute (possibly of the verifier's choice) of a user's identity, and then the rest of the proof that a user with this attribute has been vaccinated can be communicated via a zero-knowledge proof.


\section{Paper card based solution}\label{sec:method}
We use existing and well-studied cryptographic building blocks in the following protocols for performing operations such as $\sign$ and $\verify$ which can be built using a commonly used digital signature scheme, one such example is ECDSA~\cite{boneh2015graduate}.\\
In the paper-based solution we focus on keeping the proposed system as inter-operable as possible with the existing system. The user receives a coupon from the \textit{distributor} (e.g. city) and goes to the \textit{pharmacy} for getting vaccinated. Upon vaccination, the pharmacy provides the user a QR code that contains information related to the vaccination status, we refer this piece of information as \textit{badge}. The \textit{badge} can be shown by the user for proving their vaccination status. Now we describe the protocol in more detail -
\subsection{Protocol}
\paragraph{Coupon distribution}
The coupon code is generated by the \textit{issuer} and delivered to the \textit{user} through the \textit{distributor}. The process involves the \textit{issuer} setting up a secret and verifying key for generating the coupons such that the validity of the coupon can be examined by only knowing the verifying key.
\begin{itemize}
    \item The \textit{Issuer} $I$ (e.g. CDC) first generates its key pair $(\mathit{sk}_I,\mathit{vk}_I)$ for the signature scheme.  By $\sign_{I}(m)$ we will denote the output of the digital signature signing algorithm (such as the ECDSA algorithm~\cite{boneh2015graduate}) with input as the Issuer's secret signing key $\mathit{sk}$ and the message $m$.  By $\verify_I(m,\sigma)$ we will denote the output of the corresponding (e.g., ECDSA) signature verification algorithm under the verification key $\mathit{vk}_I$ on input the message $m$ and a purported signature $\sigma$. 
    \item The \textit{issuer} then generates $n$ coupons for a given zip code and job type of the users by signing the tuple $m_i = (i, \mbox{zip code}, \mbox{job type})$ for every i'th coupon. A given coupon $c$ can be represented as $$c_i = \{m_i, \sign_I(m_i)\}$$ Finally, the set of coupons $\mathcal{C}=\{c_i|\forall i \leq n\}$ is given to a local \textit{distributor} based on zip code and job type.
    \item The \textit{distributor} gives the coupon to the users after performing an eligibility check.
\end{itemize}

\begin{figure}[h]
    \centering
    \includegraphics[width=.7\columnwidth]{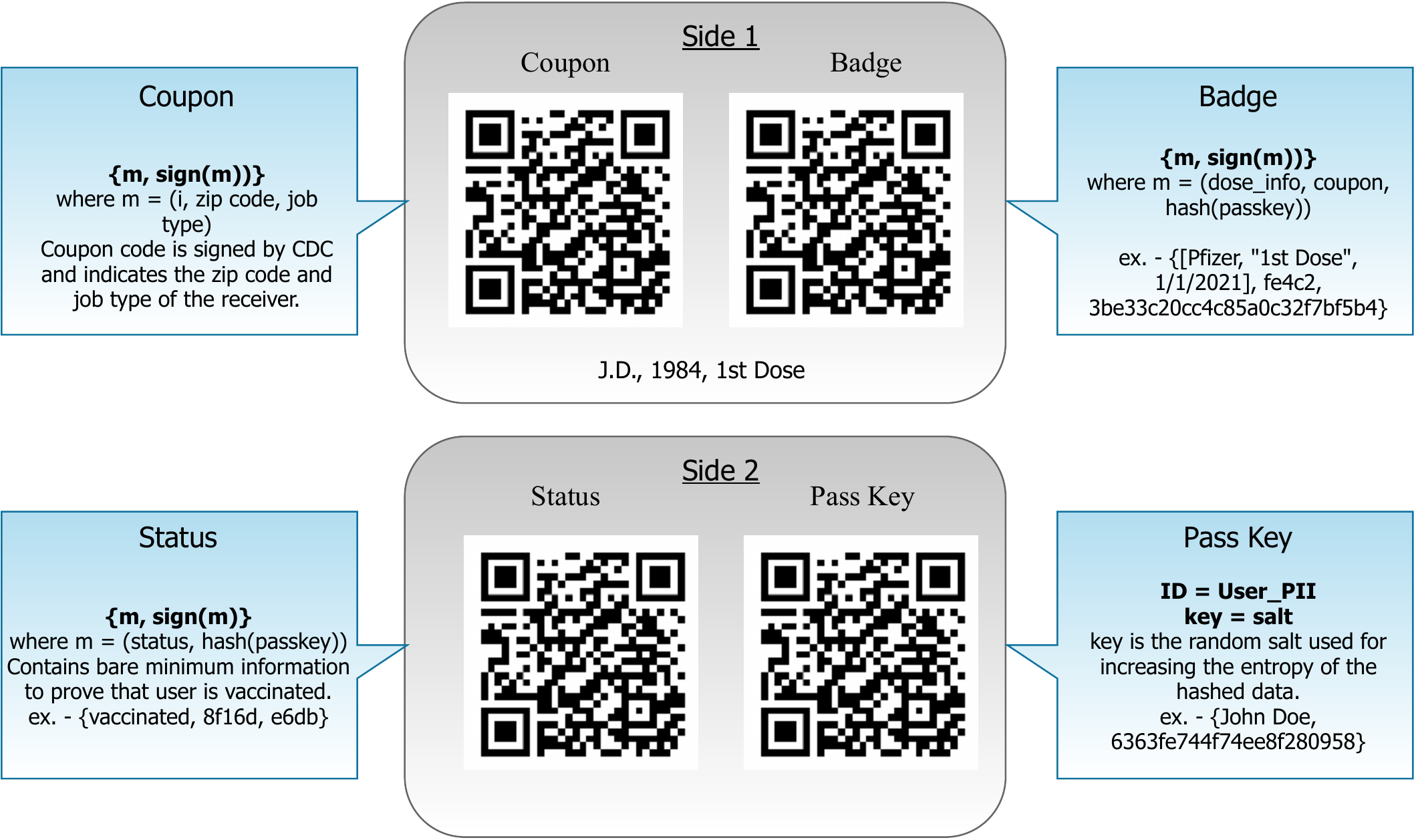}
    \caption{The representation of the paper based solution and how the QR code representation of different stickers could look like. QR code 1 carries coupon information that is used by the user for getting vaccination, QR code 2 carries badge information that consists of information related to vaccination and tied to user identity using passkey. QR code 3 has proof of vaccination and the QR code 4 has passkey of the user that contains PII.}
    \label{fig:qrcodes}
\end{figure}

\paragraph{Vaccination} In this phase, the user gets vaccinated and receives three QR-codes in addition to the previously received coupon code. As shown in Figure~\ref{fig:qrcodes}, all four QR-codes are printed on the two sides of the paper with the first side consisting of coupon and badge, and the other side has the status and pass-key. The high level idea is to keep detailed information about the user's vaccination on the first side and keep the minimal information required for proving the vaccination status on the other side.
\begin{itemize}
    \item The \textit{user} gets his coupon $c$ and shows it to the \textit{pharmacy} to get vaccinated.
    \item The \textit{pharmacy} runs $\verify_I(c)$ to validate the coupon $c$ and then checks whether the coupon has been used up already using VAMS-like system\footnote{\url{https://www.cdc.gov/vaccines/covid-19/reporting/vams/index.html}}. Upon successful validation, \textit{pharmacy} vaccinates the user.
    \item The pharmacy collects the user's PII and verifies that it matches the user's ID (e.g. - by looking at the user's driving license). User\_PII refers to the information tied to the identity of the \textit{user} (e.g. - user's full name, date of birth and etc.), which would be used for vaccination status verification in the future.
    \item The \textit{pharmacy} generates a header $h$ such that $h=\{\text{dose\_info}, c\}$ and generates $$\text{badge\_info}=\{\text{dose\_info}, c, \hash(\text{User\_PII, salt})\}$$ and submits it to the \textit{issuer} for signing it.  Salt refers to a high entropy number to ensure sufficient randomness of the hashed value; in particular, this ensures that the hashed value itself does not reveal a user's PII.
    \item The \textit{badge issuer} (that could be the same issuer as the entity that issues the coupons or a different entity with a different signing key) receives the \textit{pharmacy} request for signing the badge\_info. The \textit{issuer} updates its publicly visible registry by marking the coupon $c$ as used and signs the digest of the user information by performing two signatures $\sign_I(\text{badge\_info})$ and $\sign_I(\text{\{vaccinated, $\hash$(passkey)\}})$ and returns it back to \textit{pharmacy}. Here vaccinated refers to the minimal information required to indicate a user is vaccinated. This could be as minimal as 0 (not vaccinated) and 1 (vaccinated). However, the standards might require indicating relatively more information such as second dose, date of vaccination and etc.
    \item The \textit{pharmacy} issues two QR-codes for the user - $$\textit{badge} = \{\text{badge\_info}, \sign_I(\text{badge\_info})\}$$
    $$\textit{status} = \{\text{\{vaccinated, $\hash(passkey)$\}}, \sign_I(\text{\{vaccinated, $\hash(passkey)$\}})\}$$
    $$\textit{passkey} = \{\text{User\_PII, salt}\}$$
    \item The \textit{user} has finally four QR-codes with them \textit{coupon}, \textit{badge}, \textit{status}, and \textit{passkey}.
    \item \textbf{Second dose} The \textit{user} can receive the second dose in the future from any of the \textit{pharmacy} by presenting their \textit{badge} which would be verified by the \textit{pharmacy} first and the vaccination procedure would the proceed as above. The granular information about the vaccination enables the \textit{pharmacy} to administer the right second dose of vaccine for the \textit{user}.
\end{itemize}

\paragraph{Verification} The verification process happens in two phases - first, the validity of the badge is examined to make sure the \textit{user} is carrying a valid vaccination proof and then the identity of the \textit{user} is examined to make sure that the QR-codes belong to the person carying it.
\begin{itemize}
    \item The \textit{user} goes to the \textit{venue} and present their badge.
    \item The \textit{venue} verifies the integrity and validity of the badge by performing $\verify$(badge) 
    \item The \textit{venue} request the pass-key if they want to verify the identity of the \textit{user}. The identity is verified using a government issued ID. \textit{User} may or may not give the pass-key based on their preference.
\end{itemize}

\subsection{Security}\label{sec:sec_paper}
There are security benefits of fragmenting the user information into three distinct QR-codes. First, User\_PII does not get centralized at any point under the assumption that \textit{pharmacy} and \textit{venues} do not collude with CDC or with some global data warehouse. The protocol prevents \textit{users} from forging the vaccination proof by using digital signatures. However, a \textit{user} can sell their coupon to other users in the same zip code. While using zip\_code and job\_type impedes the potential buyers, more personalization could be done for the coupons. 

The proposed solution poses some privacy risks such as \textit{venues} keeping track of every user's passkey and creating a PII database of vaccinated users by coordinating with other \textit{venues}. This can be prevented by mandating the \textit{venue} owners to only perform the scanning using a trusted app that deletes the user record after the verification process is over. However, such a regulation enforced privacy would not protect against worst case attackers. Another potential privacy leakage could happen if the \textit{pharmacy} does not delete their record locally after generating the passkey. This would enable the semi-honest \textit{pharmacy} to keep a local database of private information of every \textit{user}. Pharmacies maybe required to keep the PII records in case the user loses the vaccination card and the salt values.

\begin{description}
\item[Unforgeability of coupons] This property is satisfied in the current scheme through the use of digital signatures. Under the assumption that the \textit{issuer} keeps the secret signing key (\textit{sk\_I}) secure with them, it would be computationally hard for any attacker to generate a coupon with valid signature.

\item[(Optional) Non-transferability of coupons] The coupon is personalized to the granularity of zip\_code and job\_type. Therefore, the coupon is transferable across individuals sharing a common zip\_code and job\_type.

\item[Unforgeability and non-transferability of badges] The proposed scheme provides unforgeability of the badges by the usage of digital signature where signing is performed by the \textit{issuer}. Hence, as long as the signing key (\textit{sk\_I}) is secretly kept by the \textit{issuer} the scheme should be unforgeable from a practical standpoint. The badges are also non-transferable due to the hash of User\_PII embedded inside it. This User\_PII is personally identifiable and hence no other user should share the same PII and hence it should be non-transferable.

\item[(Trade-off) Off-line phases] The proposed system is offline for the coupon distribution phase and verification phase. However, both vaccination phase require the \textit{pharmacy} to be online and connected with internet in order to communication with the \textit{issuer} for the badge generation process.

\item[User privacy from the issuer] Under the assumption of pre-image resistance of the hashing function used when generating the badge, the \textit{issuer} should not be able to infer the User\_PII by receiving its hash.

\item[(Trade-off) User privacy from the verifiers] The \textit{verifier} does not learn anything more than the vaccination status of the \textit{user}, however, it checks the government issued ID which belongs to the \textit{user} and hence this protocol does not provide full privacy protection against the \textit{verifier}.

\item[(Trade-off) Non-traceability] The current protocol does not provide protection against traceability across the venues because \textit{verifier} can store User\_PII shown as part of the verification process.

\item[Accessibility] The proposed scheme is highly accessible because it does not require any \textit{user} to own a smartphone or any other costly device.

\item[Usability] From a usability standpoint, the scheme only requires the \textit{users} to obtain a piece of paper from a \textit{distributor} and show it to the \textit{pharmacy}. From the perspective of the \textit{issuer}, they are required to maintain a secure server that has capabilities of maintaining a key infrastructure and serve the requests of the \textit{pharmacy}. The scheme requires a mechanism for the \textit{pharmacy} to be able to interact with the \textit{issuer} for issuing the badge to a \textit{user}. Similarly, \textit{verifier} requires a system that can use the \textit{issuer} public key and perform successful validation of the signatures. Barring the one-time software installation part, the remaining system can work in automated fashion and mostly requires scanning of QR-codes by the \textit{pharmacy} and \textit{verifiers}.

\end{description}

\section{App based solution}
As discussed in the Section~\ref{sec:sec_paper} the paper based solution introduces some of the security and feasibility constraints which limits the effectiveness of the overall system. We propose the following smartphone based protocol to circumvent some of the issues.
\subsection{Protocol}
\paragraph{Coupon distribution} The coupon distribution happens in a similar way described in the paper-based solution. The coupon code is generated by the \textit{issuer} and delivered to the \textit{user} through the \textit{distributor}. Unlike the paper-based solution, the \textit{user} also generates a public-private key pair and a private data structure for their personal information that would be useful for the upcoming phases.
\begin{itemize}
    \item The \textit{issuer} generates key pairs for the signature scheme and issues the coupon in precisely the same way as discussed in the paper-based solution.
    \item The \textit{distributor} provides coupon code $$c = \{(i, \text{zip code, job type}), \sign(i, \text{zip code, job type})\}$$ to the users which either is scanned in person by the \textit{user} or gets downloaded in the \textit{user} app through a confidential URL provided by the \textit{distributor}.
    \item The \textit{user} generates a public-private key pair $(pk_U, sk_U)$ such that private key $sk_U$ is stored with the user in a secure enclave~\cite{ios_secure_enclave} or sufficiently secure system where even the \textit{user} can not see it themselves but only decrypt information by using decryption API with the operating system (this property would be useful later on to prevent forgeability of the vaccination status). The \textit{user} also generates $$\text{pii\_hashes} = \hashtree(\{\text{User\_PII}_i, \text{salt}_i\big|\forall i \in \{\text{User\_PII}\}\})\}$$
    Here $\hashtree$ takes a set of \textit{user's} PII (e.g. - name, age, residence and etc.) and processes individual elements of the set to construct a merkle tree~\cite{boneh2015graduate}.
    \item \textbf{Second dose} The \textit{user} can receive the second dose reminder through the app in the future and they can go the \textit{pharmacy} and get their \textit{badge} scanned. The \textit{pharmacy} validates the \textit{badge} and the vaccination procedure would the proceed as mentioned above for the first dose.
\end{itemize}

\paragraph{Vaccination} The high level idea here is that the \textit{user} presents the coupon code at the \textit{pharmacy} then \textit{user} gets vaccinated and then \textit{user} presents their PII to the \textit{pharmacy} so that it can be signed securely by the \textit{issuer} for later use.
\begin{itemize}
    \item The \textit{user} shows their coupon code in the app for scanning at the \textit{pharmacy}.
    \item The \textit{pharmacy} validates the scanned coupon code $c$ and proceeds to vaccination upon successful validation of the coupon.
    \item After vaccinating the user, the \textit{pharmacy} generates $$\text{badge\_info}=\{\text{dose\_info}, c, \mathsf{pii\_hashes}\}$$ $$\text{v\_status} = \{\text{vaccinated}, pk_U, \mathsf{pii\_hashes}\}$$  Using the \text{pii\_hashes}, \textit{user} can perform selective disclosure of the information and still be able to prove the integrity of their badge.
    \item The \textit{pharmacy} sends $h$ to be signed by the \textit{issuer} and then returns the $badge$ and $status$ to the user by providing \{badge\_info,$\sign$(badge\_info)\} and \{v\_status, $\sign$(v\_status)\} respectively.
\end{itemize}

\paragraph{Verification} In the verification phase, the goal of the user is to prove that they have been vaccinated by providing a digitally signed information that consists of sufficient and minimal information about the \textit{user's} vaccination status and identity. In comparison to the paper based protocol, the \textit{user} can now selectively disclose only a subset of information by using the $\mathsf{pii\_hashes}$.
\begin{itemize}
    \item The \textit{user} walks in to the public venue and presents their badge.
    \item The \textit{venue} scans the badge and validates the signature by performing $\verify_I(status)$ and requests particular personal information.
    \item The \textit{user} gives information by performing consent based \textit{selective disclosure} of their private information.
    \item The \textit{venue} confirms the identity of the user based on the information received and the government issued ID.
\end{itemize}

\subsection{Security}
This protocol has all of the properties discussed in the paper-based protocol in the section~\ref{sec:sec_paper} and further improves the security by introducing the idea of selective disclosure of PII. This enables a fine grained control on the personal information for an \textit{user}. However, the previously discussed attack where the \textit{venues} could create a database of User\_PII is still possible. To circumvent such an attack, the app could provide anonymous credentials, which is the current direction of our ongoing work.

\begin{description}
\item[Unforgeability of coupons] Similar to the paper based scheme, this current scheme protects against unforgeability through the use of digital signatures. Under the assumption that the \textit{issuer} keeps the secret signing key (\textit{sk\_I}) secure with them, it would be computationally hard for any attacker to generate a coupon with valid signature.

\item[(Optional) Non-transferability of coupons] Similar to the paper based scheme, the coupon is personalized to the granularity of zip\_code and job\_type. Therefore, the coupon is transferable across individuals sharing a common zip\_code and job\_type.

\item[Unforgeability and non-transferability of badges] Similar to the paper based scheme, the proposed scheme provides unforgeability of the badges by the usage of digital signature where signing is performed by the \textit{issuer}. Hence, as long as the signing key (\textit{sk\_I}) is secretly kept by the \textit{issuer} the scheme should be unforgeable from a practical standpoint. The badges are also non-transferable due to the hash of User\_PII embedded inside it. This User\_PII is personally identifiable and hence no other user should share the same PII and hence it should be non-transferable.

\item[(Trade-off) Off-line phases] Similar to the paper based scheme, the proposed system is offline for the coupon distribution phase and verification phase. However, both vaccination phase require the \textit{pharmacy} to be online and connected with internet in order to communication with the \textit{issuer} for the badge generation process.

\item[User privacy from the issuer] Similar to the paper based scheme, this scheme provides protects the privacy from its leakage to the \textit{issuer}. Under the assumption of pre-image resistance of the hashing function used when generating the badge, the \textit{issuer} should not be able to infer the User\_PII by receiving its hash.

\item[(Trade-off) User privacy from the verifiers] In contrast to the paper based scheme, the proposed scheme allows selective disclosure to the \textit{verifier}. This prevents the \textit{verifier} from obtaining more private information than what is needed.

\item[(Trade-off) Non-traceability] Similar to the paper based scheme, this scheme is also traceable across different \textit{verifiers}. However, we talk about the extension in the next section that enables non-traceable verification.

\item[Accessibility] The system requires \textit{users} to own a smartphone and download the app. This reduces the accessibility across the population that does not have smartphone.

\item[Usability] Similar to the paper based scheme, this system only requires the \textit{users} to obtain an app and coupon from a \textit{distributor} and show it to the \textit{pharmacy}. From the perspective of the \textit{issuer}, they are required to maintain a secure server that has capabilities of maintaining a key infrastructure and serve the requests of the \textit{pharmacy}. The scheme requires a mechanism for the \textit{pharmacy} to be able to interact with the \textit{issuer} for issuing the badge to a \textit{user}. Similarly, \textit{verifier} requires a system that can use the \textit{issuer} public key and perform successful validation of the signatures. Barring the one-time software installation part, the remaining system can work in automated fashion and mostly requires scanning of QR-codes by the \textit{pharmacy} and \textit{verifiers}. In contrast to the paper based scheme, carrying a smartphone is more convenient than a paper based credential, making this scheme more usable.
\end{description}


\subsection{Non-traceable verification}
The proposed protocol has the limitation of being traceable across venues. We propose to fix the issue by integrating anonymous credentials into the verification protocol~\cite{camenisch2001efficient}. The anonymous credentials module can be implemented efficiently using the recent efficient versions of anonymous credentials~\cite{baldimtsi2013anonymous}.
The user would use different attributes for obtaining their credentials. During the verification phase, the verifier would request a random subset of attributes from the user for which the user has to present a physical proof. Post submission, the user would \textit{proove} the signed message by performing the verification phase of anonymous credentials~\cite{baldimtsi2013anonymous}.

\section{Contact-less Group Verification}

\begin{wrapfigure}{r}{0.5\textwidth}
\vspace{-7mm}
\centering
    \includegraphics[width=.5\columnwidth]{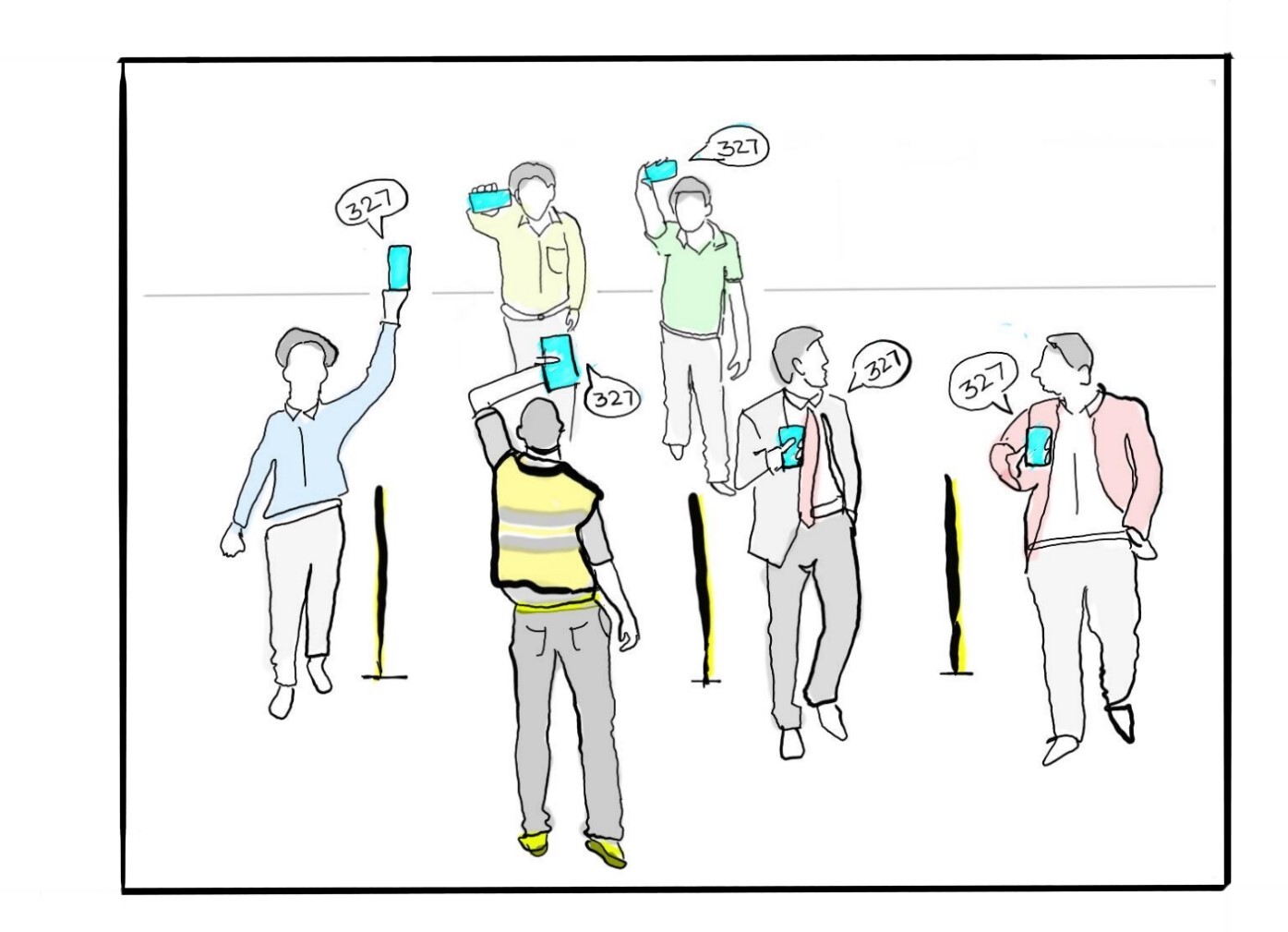}
    \caption{Depiction of the contactless group protocol where individuals perform information exchange over a wireless channel securely and generate a code on their smartphone app that proves their vaccination status to get access.}
    \label{fig:group_verification}
    \vspace{-16mm}
\end{wrapfigure}

    The high-level idea of the contact-less group verification is that the user carries a smartphone with them and the \textit{verifier} shares a secret number on the wireless channel such that only the users with correct vaccination status can present this secret number on their screen and walk-in. Figure~\ref{fig:group_verification} depicts the group verification protocol where individuals display the generated code to the guard.\\
    The contact-less group verification uses a secure channel for communication between every \textit{user} and the \textit{verifier} to prevent against any spoofing based threat that is possible in a wireless communication. The secure channel can be established using the TLS protocol, for more details about the design and implementation of the protocol, we refer the reader to the RFC~\cite{rfc8446}. To establish the TLS connection, the \textit{verifier} can share the certificate in three possible ways -\\
    \begin{itemize}
        \item Certificate signed by the \textit{issuer} - The \textit{issuer} can provide digital certificate for the public key of the venue and let \textit{user} exchange keys with them.
        \item Self signing certificate by scanning venue QR code - The \textit{user} can scan QR code available near the venue to obtain its digital certificate and trust it for further establishing the secure channel.
        \item Self signing certificate using key exchange - The \textit{venue} can display a code that can be entered by the \textit{user} in their app to trust a given certificate being broadcasted wirelessly. This code is tied to the public key used in the a digital certificate produced by the \textit{verifier}.
    \end{itemize}
    In the following we describe an interactive protocol that occurs wirelessly using channels like Bluetooth, NFC and other commonly available sensors on smartphones. The \textit{verifier} system generates unique and unguessable $k$ repeatedly for small time windows. Every \textit{user} performs one message exchange with the \textit{verifier} system and after this message exchange, everyone should have a value $k$ on their device that they can show to obtain access. Here $k$ can be a number, color, or an image the user will eventually show to the verifier visually. The challenge $k$ can change continuously. For example, a guard at the venue can change $k$ every minute.
    \begin{itemize}
    \item The \textit{user} first establishes a TLS connection with the \textit{verifier}
    \item The \textit{user} then sends $\mathsf{status}$ to the \textit{verifier} on the secure channel.
    \item The \textit{verifier} validates the message by performing $\verify_I(\text{status})$ to the user.
    \item The \textit{verifier} responds with the challenge $k$ by sending $c={\Encrypt(k, pk_U)}$
    \item The \textit{user} decrypts the packet by $k=\Decrypt(c, \sk_i)$.
    \item The \textit{user} then shows the value $k$ on the phone to the verifier.
\end{itemize}


\section{Assessing Health Outcomes of Safety and Efficacy}

The health outcome assessment requires monitoring and reporting from the user standpoint as well as a vaccine provider standpoint. In the following section, we describe how we merge the bottom-up and top-down approaches of health outcome assessment being performed.
\paragraph{Upload:} Users can upload symptoms data at any point using their coupon code ($c$). The \textit{issuer} can validate the upload using the coupon number and associate it with a verified vaccination by the \textit{pharmacy}. \textit{User} can also ask their doctor to submit an adverse reaction report using the same $c$. The upload of symptoms can be made privacy preserving by aggregating data using multi-party computation based aggregation methods~\cite{10.1145/2516930.2516944,10.5555/3154630.3154652} on top of which differentially private mechanisms~\cite{dwork2016calibrating,dwork2008differential} can be used to ensure privacy of individuals over the aggregate statistics.
\paragraph{Download:} How can the  user get an alert if their dosage batch is faulty? Or users with specific health conditions maybe at risk? We want to achieve this without the need for the user to reveal everything about themselves.
\begin{wrapfigure}{r}{0.5\textwidth}
\vspace{-7mm}
    \centering
    \includegraphics[width=0.5\columnwidth]{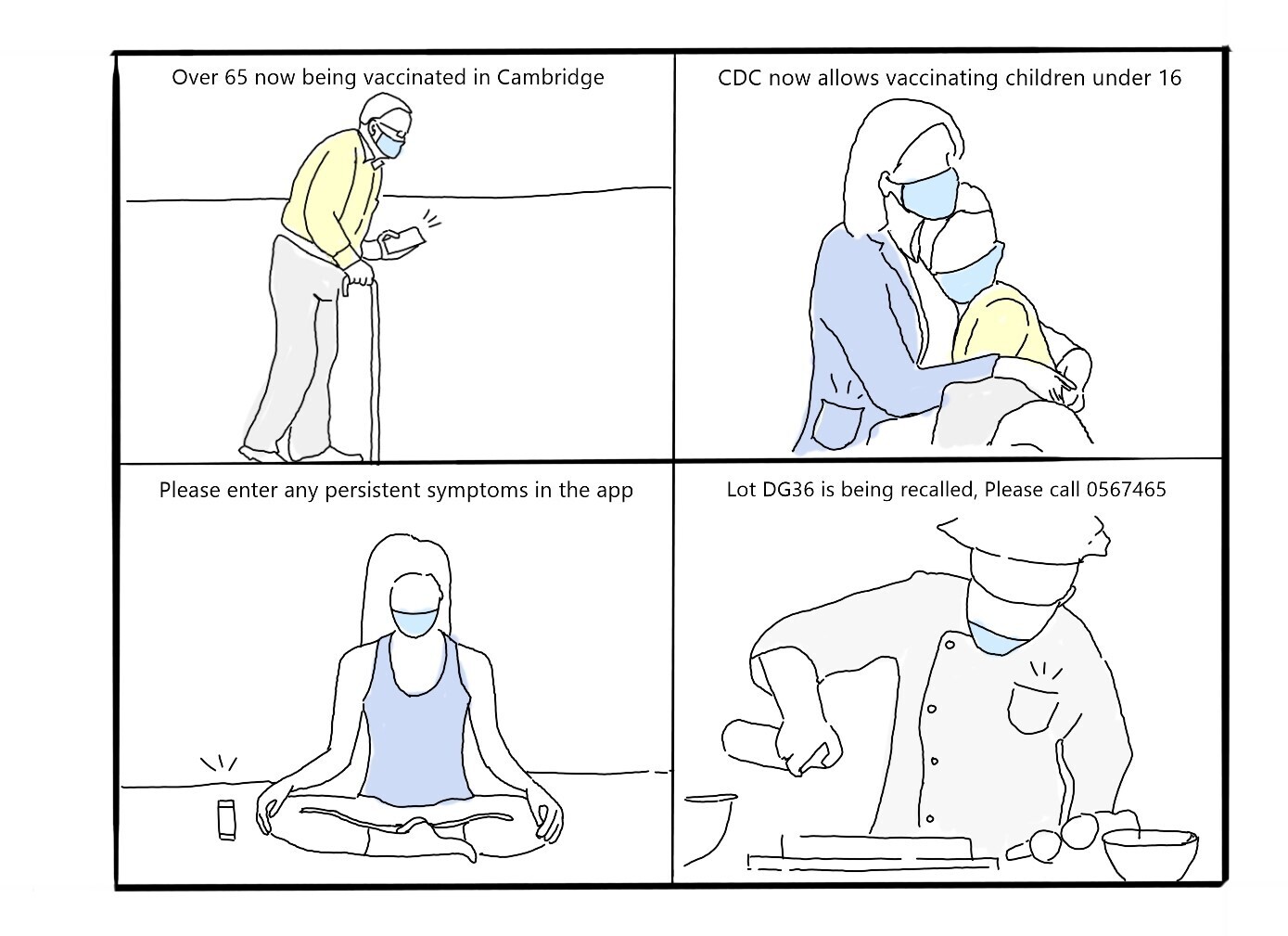}
    \caption{Health reporting and vaccination monitoring system would be used for multiple scenarios including alerts, symptom check-in, vaccine recall and etc. as shown in the figure.}
    \label{fig:health_reporting}
\vspace{-12mm}
\end{wrapfigure}
Similar to GAEN~\cite{gaen} key server, the app can download the \textit{adverse events data report} that is public for their state, every morning. The app checks if their own dose batch (company, batch or vaccination site) has any public alerts. The app also checks if there is a specific alert for their health condition (e.g. vaccine may have an adverse reaction to certain food allergy or immune health conditions)
\paragraph{Aggregate view:} How can a vaccine maker, a US state or the CDC have a detailed or aggregate view about the population level statistics? 
(i) With anonymity, a user can be tracked using the provided coupon all through their journey. If we do not want the user coupon to be tracked across the  journey, the user can upload the symptoms without the coupon code.
(ii) If the user does not wish to upload symptoms in the raw, we propose to use a protocol based on secure multi party computation similar to Prio~\cite{10.5555/3154630.3154652} to provide an aggregate statistic without revealing the privacy of any individual. 
(iii) For users without an app, they can log into V-SAFE~\cite{monitoringensuring} or VAERS~\cite{zhou2003surveillance} system using their coupon code. Similarly, a doctor updating adverse reaction report can use the coupon code.

\section{Limitations and Attacks}
\textbf{Trust ecosystem:} It is important to highlight the role of the entire ecosystem when the current public health emergency is over. Whether we could continue using the same system once the vaccination drives are over at a global scale would depend upon several factors like emerging variants, timescale of immunity against the virus. While there might be a possibility of using some of these systems for vaccination of other infectious diseases like flu, a careful re-assessment of the privacy-utility trade-off would be required to make sure that it is not happening under the pressure of public health emergency. The key infrastructure proposed in this paper should be dismantled in such a way that it can not be re-used for other identification process. The database held by the \textit{issuer} can be deleted once the information about vaccination would not be required anymore, however, this might happen at a different time period than key revocation since linked symptoms data with the \textit{issuer}'s vaccination database might be useful for retrospective studies.

Digital tools for pandemic response can have privacy and ethics issues at various fronts~\cite{raskar2020apps}, therefore we discuss some of the issues and potential pitfalls for the proposed technology. The proposed protocol provides anonymity but not privacy if the \textit{user} has to interact with the system. The ability to track the coupon $c_i$ for any \textit{user} $i$ provides pseudoanonymity which is not a full proof notion of privacy.

Because health services and verified access requires human interaction, we expect systems will require \textit{users} to furnish a state ID or similar equivalent. Therefore, the name or some identifying information needs to be embedded as part of the record code. While we store this personal information in a hashed signature, the system does not protect against the notion of plausible deniability. This can be solved by having the \textit{user} information stored exclusively with them while letting the \textit{user's} app to prove the correctness of the information using zero knowledge proof.

In our proposed protocol, the \textit{pharmacy} knows the information about the \textit{user}, and hence in the worst case, can collude with the \textit{issuer} to reveal personal information and identify a given \textit{user}. However, this targeted attack makes the assumption of \textit{pharmacy} storing the \textit{user} data instead of scanning and processing it on an ephemeral storage.
\section{Acknowledgements}
We would like to thank Vitor Pamplona for discussions and feedback. We would like to thank Priyanshi Katiyar for helping us with the sketches used in this paper.
\bibliographystyle{plain}
\bibliography{healthpassport}

\end{document}